\documentstyle[12pt]{article}
 
\newcommand{\nit}{\noindent}
\newcommand{\nl}{\newline}
\newcommand{\np}{\newpage}
\newcommand{\dsp}{\displaystyle}
\newcommand{\vs}[1]{\vspace{#1 ex}}
\newcommand{\bfr}{\begin{flushright}}
\newcommand{\efr}{\end{flushright}}
\newcommand{\bc}{\begin{center}}
\newcommand{\ec}{\end{center}}
\newcommand{\ben}{\begin{enumerate}}
\newcommand{\een}{\end{enumerate}}

\newcommand{\be}{\begin{equation}}
\newcommand{\ee}{\end{equation}}
\newcommand{\ba}{\begin{array}}
\newcommand{\ea}{\end{array}}
\newcommand{\ct}{\cite}
\newcommand{\bit}{\bibitem}

\newcommand{\ag}{\alpha}
\newcommand{\bg}{\beta}
\newcommand{\gam}{\gamma}
\newcommand{\del}{\delta}
\newcommand{\eps}{\epsilon}
\newcommand{\ve}{\varepsilon}
\newcommand{\zg}{\zeta}
\newcommand{\bzg}{\bar{\zeta}}
\newcommand{\thg}{\theta}
\newcommand{\kg}{\kappa}
\newcommand{\lb}{\lambda}
\newcommand{\sg}{\sigma}

\newcommand{\vf}{\varphi}

\newcommand{\og}{\omega}
\newcommand{\Gam}{\Gamma}
\newcommand{\Del}{\Delta}

\newcommand{\lh}{\left(}
\newcommand{\rh}{\right)}

\newcommand{\pl}{\partial}

\newcommand{\kgp}{\kg_+}
\newcommand{\kgt}{\kg_{\times}}

\font\tbf = cmbx12

\begin{document} 
 
\pagestyle{empty} 
 
\begin{flushright} 
NIKHEF/00-023 
 
\end{flushright} 
 
\begin{center} 
{\Large{\bf Motions and world-line deviations in}} \\ 
\vs{2} 
 
{\Large{\bf Einstein-Maxwell theory}} \\ 
\vs{5} 
 
{\large A.\ Balakin$^{*}$, J.W.\ van Holten$^{\dagger}$} \\ 
\vs{3} 
 
{\large and R.\ Kerner$^{**}$} \\ 
\vs{10} 
 
{\small {\bf Abstract}} 
\end{center} 

\noindent 
{\small 
We examine the motion of charged particles in gravitational and
electro-magnetic background fields. We study in particular the 
deviation of world lines, describing the relative acceleration 
between particles on different space-time trajectories. Two special 
cases of background fields are considered in detail: (a) pp-waves, 
a combination of gravitational and electro-magnetic polarized plane 
waves travelling in the same direction; (b) the Reissner-Nordstr{\o}m 
solution. We perform a non-trivial check by computing the precession 
of the periastron for a charged particle in the Reissner-Nordstr{\o}m 
geometry both directly by solving the geodesic equation, and using 
the world-line deviation equation. The results agree to the order of 
approximation considered. }  
\vs{2} 

\vfill 
\nit 
\footnoterule 
\nit 
{\small
$^{*}$ University of Kazan; Kazan, Tatarstan (Russia) 
\nl 
{\em e-mail:} {\tt{marina@iopc.kcn.ru}} \nl 
$^{**}$ Univ.\ Pierre et Marie Curie - CNRS ESA 7065 \nl 
{\em address:} Tour 22, Boite 142; 4 Place Jussieu, 75005 Paris, 
France \nl 
{\em e-mail:} {\tt{rk@ccr.jussieu.fr}} \nl 
${\dagger}$ NIKHEF; {\em address:} P.O.\ Box 41882, 1009 DB 
Amsterdam, the Netherlands} 
\nl 
{\em e-mail:}  {\tt{v.holten@nikhef.nl}} 
\np 

\pagestyle{plain}
\pagenumbering{arabic}

\section{Introduction} 

According to the equivalence principle, the motion of structureless 
test particles in a gravitational background field is determined 
only by  the space-time geometry: the worldlines are geodesics 
\ct{ae}. However, as only relative accelerations (tidal forces) 
have an invariant meaning in general relativity, the relevant 
information about the space-time geometry as a gravitational 
influence manifests itself in the rate at which worldlines diverge 
(or converge). The equation describing the rate of geodesic 
deviation in a quantitative way is well-known since at least half 
a century \ct{synge}, and numerous applications have been developed 
up to the present time \ct{2}-\ct{5}. 
 
Things become more complicated for test particles which are not 
structureless, but carry e.g.\ non-vanishing charge or spin. In 
such cases the world-line of the test particle in general is no 
longer a geodesic, but is modified by electro-magnetic and/or 
spin-orbit forces \ct{jw1,krip,jw2}. In these cases one also can 
write an equation for world-line deviations, describing relative 
accelerations, which are now due to the combined effect of 
gravitational and/or electromagnetic forces. Even a generalized 
string-theoretical version has recently been proposed \ct{6}.  
 
Nevertheless, in the presence of such additional forces the 
geometrical interpretation of the world-line deviation is not 
completely lost. For example, it is well known \ct{7,8}, that 
the geodesic lines in the five-dimensional Kaluza-Klein theory 
represent the worldlines of charged massive particles, moving 
under the influence of the Lorentz force in a curved 
four-dimensional space-time. In fact, the equivalence principle 
holds in five dimensions as well, although in four dimensions it 
applies separately to classes of particles characterized by the 
same value of the charge-to-mass ratio  $q/m$. Recently it has 
been checked \ct{9} that also the geodesic deviation equation in 
the five-dimensional Kaluza-Klein manifold leads to a generalized 
deviation equation for the world lines of charged particles, 
which can be obtained by a direct variation of the world line's 
equation in the four-dimensional space-time. Similarly, the 
world lines of particles with spin can be obtained from the 
supersymmetric extension of the geodesic equation for simple 
point particles \ct{jw1,jw2}. In spite of these developments, 
examples of a thorough analysis of solutions to the world-line 
deviation equations, as well as discussions of their physical 
implications, are scarce in the literature (\cite{Aliev-Galtsov}). 
 
\newpage 
\indent 
In this article we initiate a self-contained discussion of the 
world-line deviation equations in Einstein-Maxwell theory. We 
first derive the basic form of the equation, taking the equation 
of motion itself as a starting point. We also show that these 
equations can be derived from an action principle. We then turn 
to discuss two examples for which the nature of the solutions to 
the world-line deviation equation can be explicitly exhibited. 
The first case describes the effects of combined non-stationary 
gravitational and electromagnetic fields, in the form of a 
gravitational pp-wave. The second case concerns the analysis of 
deviation of world lines in stationary spherically symmetric 
field generated by an electrically charged mass 
(Reissner-Nordstr{\o}m solution). 

In part, the choice of these particular solutions is determined 
by their high symmetry properties, which make it possible to 
arrive at exact and explicit solutions for the background field, 
for the world lines, as well as for the deviation itself. But 
another reason for investigating this type of background solutions 
derives from the physical aspects: according to General Relativity 
the electromagnetic field acts on massive charged particles in two 
ways, directly via the Lorentz force it exerts, and indirectly, 
via its contribution to the ambient space-time curvature. The 
combination of these effects may lead to interesting new states 
of motion. Indeed, the Reissner-Nordstr{\o}m solution provides 
a particularly illustrative example of this situation, as the 
repulsive electric force on a test particle can be compensated 
by the attractive gravitational force. 
 
The supersymmetric extension of the models discussed here describes 
the modifications in the physics due to the presence of spin. These 
aspects will be discussed in a separate paper \ct{bhk2}.
 
\section{World-line deviation equations} 
 
Before considering the more general case, let us recall the 
derivation of the equation for geodesic deviation relevant to 
simple chargeless point particles. Consider a continuous smooth 
vector field $u^{\mu}(x)$ with the property 
\be 
u^{\nu} \nabla_{\nu} u^{\mu}\, =\, 0. 
\label{2.1} 
\ee 
A vector field of this kind, being transported parallel to 
itself, represents the field of vectors tangent to a smooth set 
of geodesic curves $x^{\mu}(\tau)$: 
\be 
u^{\mu}\, =\, \frac{dx^{\mu}}{d\tau}, 
\label{2.2} 
\ee 
where as usual for time-like geodesics we take the parameter $\tau$ 
to represent proper time. In terms of proper-time derivatives 
eq.(\ref{2.1}) then becomes
\be  
\frac{D^2 x^{\mu}}{D\tau^2}\, =\, \frac{d^2 x^{\mu}}{d\tau^2}\, +\,
  \Gam_{\lb\nu}^{\;\;\;\;\mu}\, \frac{dx^{\lb}}{d\tau}\,
   \frac{dx^{\nu}}{d\tau}\, =\, 0.
\label{2.2.1} 
\ee  
Now consider a slicing of space-time into smooth (1+1)-dimensional  
space-time planes formed by time-like geodesics, and in each of  
these planes a second smooth set of geodesic curves  
$x^{\mu}[\lb;\tau]$, connecting the points on the geodesics 
corresponding to a fixed value of the proper-time 
parameter\footnote{In order for these curves to be well-defined, 
one fixes the origin of proper time on each geodesic to be located 
on a smooth, differentiable curve cutting all the geodesics in 
the slice.} $\tau$. Thus, for each value of $\tau$ there is a 
smooth curve in the plane parametrized by the real affine parameter 
$\lb$. Let $n^{\mu}(x) = dx^{\mu}/d\lb$ be the field of tangent 
vectors to these curves. We observe that 
\be
u^{\nu} \nabla_{\nu} n^{\mu}\, =\, \frac{d^2 x^{\mu}}{d\tau d\lb}\,  
  +\, \Gam_{\nu\kg}^{\;\;\;\;\mu}\, \frac{dx^{\nu}}{d\tau}\,  
  \frac{dx^{\kg}}{d\lb}\, =\, n^{\nu} \nabla_{\nu} u^{\mu}.  
\label{2.3} 
\ee  
Combining eqs.(\ref{2.1}) and (\ref{2.3}) we then obtain 
$$u^{\lb} \nabla_{\lb}\, \lh u^{\nu} \nabla_{\nu} n^{\mu} \rh = 
u^{\lb} \nabla_{\lb}\, \lh n^{\nu} \nabla_{\nu} u^{\mu} \rh\, =\, 
\lh n^{\lb} \nabla_{\lb} u^{\nu} \rh\, \nabla_{\nu} u^{\mu}\, +\, 
u^{\lb} n^{\nu} \nabla_{\lb} \nabla_{\nu} u^{\mu} $$ 
\begin{equation} 
= n^{\lb} \nabla_{\lb}\, \lh u^{\nu} \nabla_{\nu} u^{\mu} \rh\, 
+\, u^{\lb} n^{\nu}\, \left[ \nabla_{\lb}, \nabla_{\nu} \right]\, 
u^{\mu}\, =\, R_{\lb\nu\kg}^{\;\;\;\;\;\;\mu} u^{\lb} u^{\kg} n^{\nu}.    
\label{2.4} 
\end{equation} 
In terms of derivations in parameter space (proper time 
derivatives) this equations reads: 
\be  
\frac{D^2 n^{\mu}}{D\tau^2}\, =\, R_{\lb\nu\kg}^{\;\;\;\;\;\;\mu}  
      u^{\lb} u^{\kg} n^{\nu}, 
\label{2.5} 
\ee  
which is the standard text-book result, as can be found e.g.\ in  
\ct{12,13}, stating that the rate of change of the velocity at 
which geodesics diverge, as measured along the curves 
$x^{\mu}(\lb)$ at fixed $\tau$, is proportional to the geodesic 
separation, with the proportionality factor given by the components 
of the Riemann curvature tensor in the direction of the geodesics 
and the curves of equal proper time.  
 
It is now straightforward to extend this derivation to curves 
representing the world line of charged particles. We start from 
the equation of motion for a particle of mass $m$ and charge $q$: 
\be  
\frac{D^2 x^{\mu}}{D\tau^2}\, =\, u^{\nu} \nabla_{\nu} u^{\mu}\, 
 =\, \frac{q}{m}\, F^{\mu}_{\;\;\nu} u^{\nu}.  
\label{2.6} 
\ee  
The right-hand side of this equation represents the 
electro-magnetic and Lorentz force acting on the particle. Next 
we obtain an equation for the relative acceleration of two world 
lines, as measured by the change in tangent vectors $n^{\mu} = 
dx^{\mu}/d\lb$ to curves connecting points of equal proper time 
on smooth planes of world lines. The equation, derived by a series 
of steps similar to those of (\ref{2.4}), reads: 
\be  
\frac{D^2 n^{\mu}}{D\tau^2}\, =\, R_{\lb\nu\kg}^{\;\;\;\;\;\;\mu}  
      u^{\lb} u^{\kg} n^{\nu}\, +\, \frac{q}{m}\, F^{\mu}_{\;\;\nu} 
      \frac{Dn^{\nu}}{D\tau}\, +\, \frac{q}{m}\, 
      \nabla_{\lb} F^{\mu}_{\;\;\nu}\, u^{\nu} n^{\lb}. 
\label{2.7} 
\ee 
In practice it is often simpler to work with the non-explicitly
covariant equation 
\be 
\ddot{n}^{\mu}\, +\, \lh 2 \Gam_{\kg\nu}^{\;\;\;\mu} u^{\kg} 
 - \frac{q}{m} F^{\mu}_{\;\;\nu} \rh\, \dot{n}^{\nu}\, +\, 
 \lh u^{\kg} u^{\sg} \pl_{\nu} \Gam_{\kg\sg}^{\;\;\;\mu} - 
 \frac{q}{m} u^{\kg} \pl_{\nu} F^{\mu}_{\;\;\kg} \rh\, n^{\nu}\, 
 =\, 0,
\label{nce}
\ee 
which is obtained from (\ref{2.7}) by substitution of the 
explicit expressions for the covariant derivatives and the 
Riemann curvature tensor. 

We observe, that this equation only determines the rate of 
divergence of world lines of particles with the same charge-to-mass 
ratio $q/m$. For neutral particles equation (\ref{2.5}) remains 
valid as the special case for $q = 0$. Eq.(\ref{2.7}) does not 
apply to the relative acceleration between particles of different 
charge-to-mass ratio, such as a neutral and a charged particle. 
On the other hand, it is clear that a neutral particle can feel 
the presence of an electro-magnetic field through the influence 
of its accompanying gravitational field \ct{jw3}. 
 
We also observe, that the direct influence of the electro-magnetic  
field on charged particles is strictly linear in the field strength 
$F_{\mu\nu}$, with coupling constant $q/m$. No terms quadratic  
in $F_{\mu\nu}$ are present in this equation in the context of  
classical general relativity. 

Observe, that eqs.(\ref{2.6}) and (\ref{2.7}) can be combined 
to show that $u \cdot Dn/D\tau$ is a constant on geodesics: 
\be 
\frac{d}{d\tau}\, \lh u \cdot \frac{Dn}{D\tau} \rh\, =\, 0.
\label{2.8}
\ee
Actually, this equation can be integrated to give the stronger 
condition 
\be 
u \cdot \frac{Dn}{D\tau}\, =\, 0.
\label{2.8.1}
\ee 
This follows, if we consider two worldlines $x_1^{\mu}(\tau) 
= x^{\mu}(\tau;\lb)$ and $x_2^{\mu}(\tau) = x^{\mu}(\tau;\lb + 
\Del \lb) \approx x_1^{\mu}(\tau) + \Del x^{\mu}(\tau)$, with 
\be 
\Del x^{\mu}(\tau)\, =\, n^{\mu}(\tau)\, \Del \lb. 
\label{2.8.2}
\ee 
As both $x_{(1,2)}(\tau)$ are solutions of the equations of 
motion, it follows that 
\be 
\ba{lll} 
g_{\mu\nu}(x_1)\, u^{\mu}_1 u^{\nu}_1 & = & -1\, =\, 
 g_{\mu\nu}(x_2)\, u^{\mu}_2 u^{\nu}_2 \\ 
 & \approx & \dsp{ g_{\mu\nu}(x_1)\, u^{\mu}_1 u^{\nu}_1\, 
 +\, 2 \Del \lb\, g_{\mu\nu}(x_1)\, u_1^{\mu}\, 
 \frac{Dn^{\nu}}{D\tau}. } 
\ea 
\label{2.8.3}
\ee 
As a result only three components of the rate of change of the 
deviation vector $n^{\mu}(\tau)$ are independent.
 
The equation (\ref{2.7}) for the world-line deviation of charged
particles in general relativity, and {\em per force} the geodesic
deviation equation (\ref{2.5}), can be derived from a covariant
action principle. The relevant lagrangean reads
\be 
L = \frac{1}{2}\, g_{\mu\nu} \frac{Dn^{\mu}}{D\tau}\, 
   \frac{Dn^{\nu}}{D\tau} + \frac{1}{2}\, R_{\kg\mu\lb\nu}\, 
   u^{\kg} u^{\lb} n^{\mu} n^{\nu} + \frac{q}{2m}\, 
   F_{\mu\nu} n^{\mu} \frac{Dn^{\nu}}{D\tau} + 
   \frac{q}{2m}\, \nabla_{\mu} F_{\nu\lb} u^{\lb} n^{\mu} 
   n^{\nu}. 
\label{3.1} 
\ee 
After some manipulations involving the Bianchi identity for 
the electro-magnetic field, the world-line deviation equation 
(\ref{2.7}) is now obtained by requiring the action $S = \int 
L d\tau$ to be stationary under variations of the deviation 
vector $n^{\mu}$: 
\be 
\frac{\del S}{\del n^{\mu}}\, =\, 0.
\label{3.2} 
\ee 
\noindent 
This action can serve as a starting point for a Hamiltonian 
formulation and for quantization. 

\section{Motion in $PP$-waves} 

\subsection{Particle worldlines}
 
The combined Einstein-Maxwell equations admit parallel plane-wave 
solutions, known as $pp$-waves\footnote{Some reviews can be found 
in \ct{11,12,jw3,jw4}.}. The non-singular metrics of these 
solutions are represented by the line-element 
\be 
c^2 d\tau^2\, =\, - dudv\, - \, K(u,x,y)\, du^2\, +\, dx^2\, + \, dy^2, 
\label{4.1} 
\ee 
which reduces to flat Minkowski space-time when the metric 
$uu$-component vanishes: $K(u,x,y) = 0$. As is implicit in 
(\ref{4.1}), we choose to work with the light-cone co-ordinates 
\be 
u = ct -z, \hspace{3em} v = ct + z. 
\label{4.2} 
\ee 
As the determinant of this metric is negative definite, it is 
manifestly invertible everywhere. 

The solutions for the vector potential of the electro-magnetic 
field are of the form 
\be 
A_{\mu}\, =\, \del_{\mu}^x A_x(u)\, +\, \del_{\mu}^y A_y(u). 
\label{4.3} 
\ee 
The corresponding electric and magnetic fields are wave-like, 
propagating in the $z$-direction, transverse and orthogonal: 
\be 
\frac{E_i}{c}\, =\, -\, \eps_{ij}\, B_j\, =\, A_i^{\prime}(u), 
\hspace{3em} i,j = (x,y). 
\label{4.4} 
\ee 
Here the prime denotes a derivative with respect to $u$. These 
fields satisfy the free Maxwell equations in a space-time with 
metric (\ref{4.1}). The Einstein equations, with the 
energy-momentum tensor of the electro-magnetic field, 
reduce now to the single p.d.e.\  \ct{11,jw3,jw4} 
\be 
\Del_{trans} K\, =\, \frac{16\pi \eps_0 G}{c^4}\, E_i^2. 
\label{4.5} 
\ee 
Here $\Del_{trans}$ is the ordinary 2-dimensional laplacian 
in the $x$-$y$-plane. The general solution for the metric 
component $K$ is 
\be 
K(x,y,u)\, =\, \frac{4\pi \eps_0 G}{c^4}\, E_i^2\, (x^2 + y^2)\, 
 +\, f(u,\zg) + \bar{f}(u,\bzg), 
\label{4.6} 
\ee 
with $f$, $\bar{f}$ conjugate holomorphic functions of the complex
transverse co-ordinates 
\be 
\zg = x + i y, \hspace{3em} \bzg = x - i y. 
\label{4.7} 
\ee 
However, as constant or linear functions of $(\zg, \bzg)$ do not 
give rise to a true gravitational field (the full Riemann tensor 
vanishes for them), we take $f(u,\zg)$ to be an analytic function 
of the type 
\be 
f(u,\zg)\, =\, \sum_{n=2}^\infty\, \kg_n(u) \zg^n. 
\label{4.8} 
\ee 
Because of its behaviour under rotations in the transverse plane 
we can identify the first term with $n = 2$ as free quadrupole 
gravitational waves. In contrast, the first term in (\ref{4.6}), 
which is the special solution due to the presence of the 
electro-magnetic waves, is invariant under rotations in the 
transverse plane. 
 
The equations of motion (\ref{2.6}), adapted to this case, can be 
obtained from the world-line lagrangean 
\be 
L = \frac{1}{2}\, \lh - \dot{u} \dot{v} - K \dot{u}^2 + \dot{x}^2 
    + \dot{y}^2 \rh\, -\, \frac{q}{m}\, \lh A_x \dot{x} + A_y 
    \dot{y} \rh. 
\label{4.9} 
\ee 
Three of the equations of motion can be written as 
\be 
\ba{ll} 
\dot{u}\, =\, \mbox{constant}\, \equiv\, \gam, & \\ 
 & \\
\dsp{ \ddot{x}\, =\, -\, \frac{\gam^2}{2}\, K_{,x}\, -\, 
  \frac{q\gam}{mc}\, E_x, } & \dsp{ 
      \ddot{y}\, =\, -\, \frac{\gam^2}{2}\, K_{,y}\, -\, 
  \frac{q\gam}{mc}\, E_y. } 
\ea 
\label{4.10} 
\ee 
The fourth equation, for the light-cone co-ordinate $v$, can be
replaced by the conservation of the world-line hamiltonian 
\be 
H\, =\, -\, 2 p_v p_u\, +\, 2 K p_v^2\, +\, \frac{1}{2}\, \lh 
 \Pi_x^2 + \Pi_y^2 \rh, 
\label{4.11} 
\ee 
where the (covariant) momenta are defined by 
\be 
\ba{ll} 
\dsp{ p_v\, =\, -\, \frac{1}{2}\, \dot{u}\, =\, -\, \frac{\gam}{2}, 
} & \dsp{ p_u\, =\, -\, \frac{1}{2}\, \dot{v}\, -\, K \dot{u}, }\\ 
 & \\ 
\dsp{ \Pi_x\, =\, p_x\, +\, \frac{q}{m}\, A_x\, =\, \dot{x}, }& 
\dsp{ \Pi_y\, =\, p_y\, +\, \frac{q}{m}\, A_y\, =\, \dot{y}. } 
\ea 
\label{4.12} 
\ee 
Here $(p_x,p_y)$ are the standard canonical momenta in the 
transverse plane, as used to define the Poisson brackets. It is 
convenient to use the first equation (\ref{4.10}) to change the 
world-line parameter from 
 proper time $\tau$ to light-cone time 
$u$. Furthermore, equation (\ref{4.11}) can be solved for the 
velocity component $\dot{v}$, or equivalently $v^{\prime} = \gam 
\dot{v}$: 
\be 
v^{\prime}\, =\, (x_i^{\prime})^2\, -\, K\, -\, \frac{2 H}{\gam^2}, 
\label{4.12.1} 
\ee 
the last term being a constant of integration. In terms of the 
constants $H$ and $p_v$ the dynamical problem is now reduced to 
that of solving for the motion in the transverse plane. Consider
the case of electro-magnetic and pure quadrupole free waves; 
writing $\kgp = 4\,$Re$\,\kg_2$, $\kgt = -4\,$Im$\,\kg_2$ the metric 
component is 
\be 
K\, =\, \frac{4\pi \eps_0 G}{c^4}\, E_i^2(u)\, (x^2 + y^2)\, 
 + \frac{\kgp(u)}{2}\, (x^2 - y^2)\, +\, \kgt(u)\, xy\, 
 \equiv\, \frac{1}{2}\, K_{ij} x^i x^j. 
\label{4.13}  
\ee 
This metric allows additional Killing vectors in the $x$-$y$-plane, 
making the dynamical problem fully solvable in terms of first 
integrals of motion. The additional constants of motion are 
given by 
\be 
\ba{lll} 
J & = & J_x \Pi_x\, +\, J_y \Pi_y\, +\, 2 p_v\, \lh 
  J_x^{\prime} x + J_y^{\prime} y \rh\, +\, \Del \\  
 & & \\ 
 & = & J_x \dot{x}\, -\, \dot{J}_x x\, +\, 
 J_y \dot{y}\, -\, \dot{J}_y y\, +\, \Del, 
\ea 
\label{4.14} 
\ee 
where the co-efficient functions $J_{(x,y)}(u)$ and $\Del$ are 
the solutions of the differential equations resulting from the 
requirement $\frac{d J}{d \tau} = 0$: 
\be 
\ba{lll} 
\dsp{ J_x^{\prime\prime}\, +\, \frac{1}{2}\,\lh \kgp + \frac{8\pi 
 \eps_0 G}{c^4}\, E_i^2\,\rh\, J_x\, +\, \frac{\kgt}{2}\, J_y }& 
 = & 0, \\ 
 & & \\ 
\dsp{ J_y^{\prime\prime}\, +\, \frac{1}{2}\,\lh - \kgp + \frac{8\pi 
 \eps_0 G}{c^4}\, E_i^2\,\rh\, J_y\, +\, \frac{\kgt}{2}\, J_x }& 
 = & 0, \\ 
 & & \\ 
\dsp{ \Del^{\prime}\, =\, \frac{q}{mc}\, \lh E_x J_x + E_y J_y 
      \rh.} 
 & & 
\ea 
\label{4.15} 
\ee 
These equations can be summarized in the compact form 
\be 
J_i^{\prime\prime}\, +\, \frac{1}{2}\, K_{ij}\, J_j\, =\, 
 J_i^{\prime\prime}\, -\, R_{iuju}\, J_j\, =\, 0, \hspace{2em} 
 \Del^{\prime}\, =\, \frac{q}{mc}\, E_i J_j\, =\, \frac{q}{m}\, 
 F_{ui}\, J_j. 
\label{4.15.1} 
\ee 
They admit a two-parameter set of solutions, which implies the 
existence of two linearly independent constants of motion fixing 
the values of the velocity components $\dot{x}$ and $\dot{y}$. 
\newline 
\indent 
This can be inferred from the trivial vanishing of the Wronskian 
$W(f,g) = (f'g-fg')$ of any two solutions, $J_{(1)}$ and $J_{(2)}$, 
of (\ref{4.14}), which reduces the original set of four constants 
of integration of the system (\ref{4.15}) to two constants only. 

\subsection{World-line deviation} 

We now discuss the generalized world-line deviation equation 
(\ref{2.7}) in the context of the electro-magnetic and free 
quadrupole gravitational waves described by the metric 
(\ref{4.1}), (\ref{4.13}) and the electro-magnetic vector 
potential (\ref{4.3}). 
 
The only non-vanishing components of the Riemann curvature 
tensor in this case are 
\be 
\ba{l} 
\dsp{ R_{xuxu}\, =\, -\, \frac{1}{2}\, K_{xx}\, =\, 
  - \frac{1}{2}\, \lh \kgp + 
  \frac{8\pi \ve_0 G}{c^4}\, E_i^2 \rh, }\\ 
 \\ 
\dsp{ R_{yuyu}\, =\, -\, \frac{1}{2}\, K_{yy}\, =\, 
  - \frac{1}{2}\, \lh - \kgp + 
  \frac{8\pi \ve_0 G}{c^4}\, E_i^2 \rh, }\\ 
 \\ 
\dsp{ R_{xuyu}\, =\, R_{yuxu}\, =\, -\, \frac{1}{2}\, K_{xy}\, 
 =\, - \frac{1}{2}\, \kgt, } 
\ea 
\label{5.1} 
\ee 
whilst the non-vanishing electro-magnetic field strength has 
two non-zero component 
\be 
F_{ux}\, =\, \frac{1}{c}\, E_x, \hspace{2em} 
F_{uy}\, =\, \frac{1}{c}\, E_y. 
\label{5.2} 
\ee 
It is then straightforward to write down the explicit form of 
the world-line deviation equations. For the $u$ component 
of the deviation vector it reduces to 
\be 
\frac{D^2 n^u}{D\tau^2}\, =\, \ddot{n}^u\, =\, 0. 
\label{5.3} 
\ee 
Hence the rate of change of $n^u$ is constant: $\dot{n}^u = \bg 
= \mbox{constant}$, and 
\be 
n^u(\tau)\, =\, n^u(0)\, +\, \bg \tau. 
\label{5.4} 
\ee 
The world-line deviation in the transverse $x$-$y$-plane is 
described by the equations 
\be 
\ba{lll} 
\dsp{ \frac{D^2 n^i}{D\tau^2} }& = & \dsp{ \ddot{n}^i\, +\, 
 \frac{\gam}{2}\, \frac{dK_{,i}}{d\tau}\, n^u\, +\, 
 \gam\, K_{,i} \dot{n}^u }\\ 
 & & \\ 
 & = & \dsp{ -\, \frac{\gam^2}{2}\, K_{ij} n^j\, +\, 
 \frac{\gam}{2}\, K_{ij} \dot{x}^j n^u\, -\, \frac{q\gam}{mc}\, 
 E_i^{\prime} n^u\, -\, \frac{q}{mc}\, E_i \dot{n}^u. } 
\ea 
\label{5.5.1} 
\ee 
This can be simplified, using $u$ as independend variable, 
to read 
\be 
n_i^{\prime\prime}\, =\, -\, \frac{1}{2}\, K_{ij} n_j\, 
 -\, \frac{1}{2}\, \lh K_{,i} n^u \rh^{\prime}\, -\, 
 \frac{1}{2}\, K_{,i} (n^u)^{\prime}\, +\, \frac{1}{2}\, 
 K_{ij} x_j^{\prime} n^{u}\, -\, \frac{q}{mc\gam}\, 
 \lh E_i n^{u} \rh^{\prime}. 
\label{5.6.1} 
\ee 
This is a system of coupled ordinary linear second order 
differential equations with field dependend co-efficients. 
In spite of their complicated form, the equations for the 
transverse components $n^x$ and $n^y$ can be solved in terms 
of the constants of motion (\ref{4.14}), (\ref{4.15}). First 
define the linear combinations 
\be 
N_i \equiv n_i - x_i^{\prime} n^u + x_i (n^u)^{\prime}, 
 \hspace{3em} i = (x,y). 
\label{5.8} 
\ee 
By substitution of the explicit form of the equations 
(\ref{5.6.1}) we obtain 
\be 
N_i^{\prime\prime}\, =\, -\, \frac{1}{2}\, K_{ij} N_j. 
\label{5.9} 
\ee 
We observe, that these equations are identical with 
eqs.(\ref{4.15}), (\ref{4.15.1}). Therefore they admit the 
solutions 
\be 
N_x = \ag J_x, \hspace{3em} N_y = \ag J_y, 
\label{5.10} 
\ee 
where $\ag$ is an arbitrary constant of proportionality; this 
constant serves to normalize the solutions for $n^{\mu}$, and 
can be absorbed into the parameter $\lb$ measuring the deviation 
between the world lines. Together with the solution (\ref{5.4}) 
for $n^u$ we now have a full solution for three of the four 
components $n^{\mu}$. 
 
The last remaining component is the second light-cone component 
$n^{v}$, which is subject to the equation 
\be 
\ba{lll} 
\dsp{ \frac{1}{\gam^2}\, \frac{D^2 n^v}{D\tau^2} } & = & \dsp{ 
  \left[ (n^v)^{\prime} + K^{\prime} n^u + K_{,i} n^i 
  \right]^{\prime}\, +\, K^{\prime} (n^u)^{\prime}\, +\, K_{,i} 
  n^{i\, \prime}\, +\, \frac{1}{2}\, K_{,i}^2 n^u } \\ 
 & & \\ 
 & = & \dsp{ K_{ij} x^{i\, \prime} x^{j\, \prime} n^u\, -\, 
 K_{ij} x^{i\, \prime} n^j\, -\, \frac{2q}{mc\gam}\, E_i 
 \lh n^{i\, \prime} + \frac{1}{2} K_{ij} x^j n^u \rh }\\ 
 & & \\ 
 & & \dsp{ -\, \frac{2q}{mc\gam}\, E_i^{\prime}\, x^{i\, \prime} 
 n^u. } 
\ea 
\label{5.7} 
\ee 
Upon substitution of the definitions (\ref{5.8}) once more, using 
the deviation equations (\ref{5.3}) and (\ref{5.9}), and using 
the equations of motion (\ref{4.10}) to eliminate the explicit 
dependence on the electromagnetic fields, this equation can be 
cast into the form 
\be 
\left[ \lh n^v \rh^{\prime} - 2 \lh N_i^{\prime}  x_i \rh^{\prime} 
 + \lh K - (x_i^{\prime})^2 \rh^{\prime} n^u - 
 \lh K - (x_i^{\prime})^2 \rh 
 n^{u\, \prime} \right]^{\prime}\, =\, 0. 
\label{5.11} 
\ee 
Finally equation (\ref{4.12.1}) can be substituted, which together 
with equation (\ref{5.3}) gives the final result 
\be 
\left[ n^v\, +\, 2 v (n^u)^{\prime}\, -\, v^{\prime} n^u - 
  2 N_i^{\prime} x_i \right]^{\prime\prime} =\, 0.
\label{5.12} 
\ee 
It follows that the second light-cone component of the world-line 
deviation is given by 
\be 
n^v\, +\, 2 v (n^u)^{\prime}\, -\, v^{\prime} n^u\, =\, 
 2 \ag J_i^{\prime} x_i\, +\, A\, +\, Bu, 
\label{5.13} 
\ee 
with $A$ and $B$ constants of integration to be fixed by the 
initial conditions. Observe, that starting out with $n^u = 
(n^u)^{\prime} = 0$, the right-hand side directly gives $n^v$ 
itself. 

It is well known \cite{12} that the $pp$-wave solution may be 
encoded in another frequently used co-ordinate system, in which 
the line element takes on the following form: 
\begin{equation} 
ds^{2} = dudv + g_{22}(u) (dx^2)^2 + g_{33}(u)(dx^3)^2, 
\end{equation} 
where the light-cone variables  $u$ and $v$ are defined as before. 

The space-time endowed with metric (11) admits Petrov's $G_5$ 
group of isometries \ct{10,11}. The three Killing vectors, 
generating the Abelian subgroup of $G_5$ are defined as: 
\begin{equation} 
\xi_{(V)}^\mu = \delta_V^\mu, \quad \xi_{(2)}^\mu = \delta_2^\mu, 
\quad \xi_{(3)}^\mu = \delta_3^\mu. 
\label{kv}
\end{equation} 
The first Killing vector is covariantly constant, isotropic and 
orthogonal to second and third vectors in (\ref{kv}). The Lie 
derivative of the metric along any Killing vector is equal to 
zero: 
$L_{\xi}( g_{\mu \nu}) =0$. 
\newline 
\indent 
Using these symmetries, we can explicitly integrate the world-line 
equations, and then also obtain the explicit solutions for the 
deviation $4$-vector in these co-ordinates. The details of these
calculations can be found in (\cite{balakin}).
 
\section{Reissner-Nordstr{\o}m fields} 
 
\subsection{Worldlines and integrals of motion}
The Reissner-Nordstr{\o}m solution of the Einstein-Maxwell 
system of equations describes the external gravitational and electro-magnetic 
fields of a static and spherically symmetric distribution of mass 
and charge, or ---in the case one takes it as a complete solution--- 
a static charged black hole of mass $M$ and charge $Q$. The 
gravitational field is described by the line-element (in natural 
co-ordinates with $c = 1$) 
\be 
g_{\mu\nu} dx^{\mu} dx^{\nu} = - d\tau^2 = - B(r) dt^2 + 
 \frac{1}{B(r)}\, dr^2 + r^2 \lh d\thg^2 + \sin^2 \thg\, 
 d\vf^2 \rh, 
\label{6.1} 
\ee 
with 
\be 
B(r) = 1 - \frac{2M}{r} + \frac{Q^2}{r^2} = \lh 1 - 
 \frac{M}{r}\rh^2 + \frac{Q^2 - M^2}{r^2}.  
\label{6.2} 
\ee 
The corresponding solution of Maxwell's equations in this 
space-time is 
\be 
A = A_{\mu} dx^{\mu} = - \frac{Q}{4\pi r}\, dt, \hspace{2em} 
 F = dA = \frac{Q}{4\pi r^2}\, dr \wedge dt. 
\label{6.3} 
\ee 
Details of this solution of the field equations are discussed in 
standard text books, see e.g.\ \ct{12}. In the following we 
assume $M^2 > Q^2$. 
 
We immediately proceed to the solution of the world-line 
equations for a test particle of mass $m$ and charge $q$. As the 
spherical symmetry guarantees conservation of angular momentum, 
the particle orbits are always confined to an equatorial plane, 
which we choose to be the plane $\thg = \pi/2$. The angular 
momentum $J$ is then directed along the $z$-axis. Denoting its 
magnitude per unit of mass by $\ell = J/m$, we have 
\be 
\frac{d\vf}{d\tau}\, =\, \frac{\ell}{r^2}. 
\label{6.4} 
\ee 
In addition, as the metric is static outside the horizon 
$r_+ = M + \sqrt{M^2 - Q^2}$, it allows a time-like Killing vector 
which guarantees the existence of a conserved world-line energy 
(per unit of mass $m$) $\ve$, such that 
\be 
\frac{dt}{d\tau}\, =\, \frac{ \ve - \frac{qQ}{4\pi mr}}{
 1 - \frac{2M}{r} + \frac{Q^2}{r^2}}. 
\label{6.5} 
\ee 
Finally, the equation for the radial co-ordinate $r$ can be 
integrated owing to the conservation of the world-line hamiltonian, 
i.e.\ the conservation of the absolute four-velocity: 
\be 
\lh \frac{dr}{d\tau} \rh^2\, =\, \lh \ve - \frac{qQ}{4\pi mr} 
 \rh^2\, -\, \lh 1 - \frac{2M}{r} + \frac{Q^2}{r^2} \rh \lh 1 + 
 \frac{\ell^2}{r^2} \rh. 
\label{6.7} 
\ee 
>From this we derive a simplified expression for the radial 
acceleration: 
\be 
\frac{d^2 r}{d\tau^2}\, =\, \frac{qQ}{4\pi mr^2}\, \lh 
  \ve - \frac{qQ}{4\pi mr} \rh\, -\, \frac{1}{r^3}\, \lh Mr - 
  Q^2\rh\, +\, \frac{\ell^2}{r^3}\, \lh 1 - 
  \frac{3M}{r} + \frac{2Q^2}{r^2} \rh. 
\label{6.6} 
\ee 
Eq.(\ref{6.7}) can in principle be integrated directly. However, 
to get directly an approximate solution to the equations of motion 
one can also study perturbations of special simple orbits. We 
follow
both paths and compare the results. First, we find a 
solution to the equation for bound orbits in terms of an integral, 
from which we can compute e.g.\ the periastron shift, similar to 
the case of the
 Schwarzschild solution; secondly, we can solve 
the simple case of
 circular orbits, and then study the problem of 
generic bound orbits by considering the world-line deviation 
equations for the special case of world lines close to circular 
orbits. This is the subject of the
 final section of this paper. 

For generic orbits, we first construct the orbital equation from 
the separate equations of motions for $r(\tau)$ and $\vf(\tau)$, 
eqs.(\ref{6.4}) and (\ref{6.7}): 
\be 
\ba{lll} 
\dsp{ \ell^2\, \lh \frac{d}{d\vf} \frac{1}{r} \rh^2} & = & \dsp{ 
 \ve^2\, -\, 1\, +\, \frac{(4\pi mM - \ve qQ)}{2\pi m r} }\\ 
 & & \\ 
 & & \dsp{ -\, \frac{1}{r^2}\, \lh \ell^2 + Q^2 - 
 \lh \frac{qQ}{4\pi m} \rh^2 \rh\, +\, \frac{2M\ell^2}{r^3}\, -\, 
 \frac{\ell^2 Q^2}{r^4}. } 
\ea 
\label{6.7.1} 
\ee 
We parametrize the solutions of this equation for $\ell \neq 0$ 
as quasi-Kepler orbits \ct{jw2}: 
\be 
r(\vf)\, =\, \frac{r_0}{1 + e \cos y(\vf)}, 
\label{6.7.2} 
\ee 
with $e < 1$ for bound states. The function $y(\vf)$, which is 
linear for Kepler orbits, here remains to be determined. As the 
extrema of the orbit (the apo- and periastron) are reached for 
$y(\vf) = (2n + 1) \pi$ (apastron) and $y(\vf) = 2\pi n$ 
(periastron) eq.(\ref{6.7.1}) evaluated at these extrema give 
two independent equations relating $e$ and 
 $r_0$ to the other 
parameters: 
\be 
\ba{lll} 
\ve^2 - 1 & = & \dsp{ - \frac{(4\pi mM - \ve qQ)}{2\pi mr_0}\, +\, 
 \frac{1}{r_0^2}\, \lh \ell^2 + Q^2 - \lh \frac{qQ}{4\pi m}\rh^2 
 \rh\, 
 \lh 1 + e^2 \rh }\\ 
 & & \\ 
 & & \dsp{ -\, \frac{2M \ell^2}{r_0^3}\, \lh 1 + 3e^2 \rh\, +\, 
  \frac{\ell^2 Q^2}{r_0^4}\, \lh 1 + 6 e^2 + e^4 \rh, } 
\ea 
\label{6.7.3} 
\ee 
and 
\be 
\ba{lll} 
\dsp{ \frac{(4\pi mM - \ve qQ)}{2\pi mr_0} } & = & \dsp{ 
 \frac{2}{r_0^2}\, \lh \ell^2 + Q^2 - \lh \frac{qQ}{4\pi m} \rh^2
 \rh\,  -\, \frac{2M \ell^2}{r_0^3}\, \lh 3 + e^2 \rh\, }\\ 
 & & \\ 
 & & \dsp{ +\, \frac{4 \ell^2 Q^2}{r_0^4}\, \lh 1 + e^2 \rh. } 
\ea 
\label{6.7.4} 
\ee 
With the help of these relations we find for the function $y(\vf)$ 
the first-order differential equation 
\be
\ba{lll} 
\dsp{ \lh \frac{dy}{d\vf} \rh^2 }& = & \dsp{ 1\, +\, 
 \frac{Q^2}{\ell^2}\, \left[ 1 - \lh \frac{q}{4\pi m}\rh^2 \right]\, 
 -\, \frac{6M}{r_0}\, +\, \frac{Q^2}{r_0^2}\, \lh 6 + e^2 \rh }\\ 
 & & \\ 
 & & \dsp{ -\, \frac{2e}{r_0}\, \lh M - \frac{2Q^2}{r_0} \rh\, 
     \cos y\, 
     +\, \frac{e^2 Q^2}{r_0^2}\, \cos^2 y. } \\ 
 & & \\ 
 & \equiv & \dsp{ A + B \cos y + C \cos^2 y } 
\ea 
\label{6.7.5} 
\ee 
It follows that the total change in the orbital angle $\vf$ between 
two periastrons is given by 
\be 
\Del \vf\, =\, \int_0^{2\pi}\, \frac{dy}{\sqrt{ A + B \cos y + 
   C \cos^2 y}}. 
\label{6.7.6} 
\ee 
As a result we obtain an expression for the periastron shift
per one revolution:
\be 
\del \vf\, =\, \Del \vf\, -\, 2\pi\, \approx\, 
 2\pi\, \lh \frac{3M}{r_0} - \frac{Q^2}{2Mr_0} \rh\, +\, 
 {\cal O}\lh e^2,\frac{M^2}{r_0^2},\frac{Q^2}{r_0^2}\rh. 
\label{6.7.7}
\ee 

\subsection{World-line deviation near circular orbits} 
 
We observe that for circular orbits $r = R =$ constant, the 
expressions for $dr/d\tau$, eq.(\ref{6.7}), and $d^2r/d\tau^2$, 
eq.(\ref{6.6}), must both vanish at all times. This produces two 
relations between the three dynamical quantities $(R,\ve,\ell)$, 
showing that the circular orbits are characterized completely by 
specifying either the radial co-ordinate, or the energy, or the 
angular momentum of the test particle. In particular, the equation 
for constant radial velocity gives: 
\be 
 \lh \ve - \frac{qQ}{4\pi mR} \rh^2\, =\, 
 \lh 1 - \frac{2M}{R} + \frac{Q^2}{R^2} \rh \lh 1 + 
 \frac{\ell^2}{R^2} \rh. 
\label{6.9} 
\ee 
Substitution of this result into the expression for the radial 
acceleration, multiplying by $R^4$, and equating the result to 
zero, leads to :
\be 
\ba{l} 
\dsp{ \left[ \frac{\ell^2}{R}\, -\, M\, \lh 1 + \frac{3\ell^2}{R^2} 
\rh\, +\, \frac{Q^2}{R}\, \lh 1 + \frac{2\ell^2}{R^2} \rh \right]^2 
}\\ 
 \\ 
\dsp{ ~~~~~~~~~=\, 
 \lh \frac{qQ}{4\pi m} \rh^2\, \lh 1 + \frac{\ell^2}{R^2} \rh\, 
 \lh 1\, -\, \frac{2M}{R}\, +\, \frac{Q^2}{R^2} \rh.}
\ea 
\label{6.10} 
\ee 
Note that the right-hand side vanishes for chargeless test 
particles, i.e.\ on geodesics. In the limiting case $Q = 0$ we 
reproduce the well-known result 
\be 
MR^2 - \ell^2 ( R - 3M ) = 0 \hspace{1em} \Rightarrow 
 \hspace{1em} R\, =\, \frac{\ell^2}{2M}\, \lh 1 + \sqrt{ 
 \dsp{ 1 - \frac{12M^2}{\ell^2} }} \rh, 
\label{6.8} 
\ee 
leading to the requirement $R \geq 6M$ for stable circular orbits 
to exist \ct{12}. 
 
Because of the spherical symmetry and the conservation of 
angular momentum {\em all} test-particle orbits are planar. 
Therefore the deviation of any bound orbit from a circular one 
can always be computed with reference to a circular orbit in 
the same plane, which may be chosen to be the equatorial plane 
$\thg = \pi/2$. For such world-line deviations the component 
of the deviation out of the plane is always zero: 
$n^{\thg}(\tau) = 0$. 

On the other hand, one might also be interested in the motion 
of a cloud of testparticles, as for example in disks of matter 
surrounding stars or planets, whose orbits are close but not 
necessarily precisely in the same plane. In such a case the 
deviation between the planes of the orbits, parametrized by 
$n^{\thg}$, oscillates between positive and negative values. 
 
In order to analyse the deviations quantitatively, we use the connection 
coefficients and curvature components as listed in the appendix, 
together with the expression (\ref{6.3}) for the electro-magnetic 
field strength, to study the world-line deviation as given by 
eq.(\ref{2.7}), with reference to a nearby circular orbit of 
radius $r = R$. We begin with the equation for the component 
$n^{\thg}$ out of the equatorial plane. Using the properties 
of the circular reference orbit: 
\be 
\ba{ll}
r = R = constant, & u^r = 0, \\ 
 & \\ 
\dsp{ \thg = \frac{\pi}{2}, }& u^{\thg} = 0, 
\ea 
\label{n.1} 
\ee 
whilst $F^{\thg}_{\;\;\nu}  = 0$, one finds from the deviation 
equation in the form (\ref{nce})  
\be 
\ddot{n}^{\thg}\, +\, (u^{\vf})^2\, \pl_{\thg} 
 \Gam_{\vf\vf}^{\;\;\;\thg}\, n^{\thg}\, =\, 
\ddot{n}^{\thg}\, +\, \og^2 n^{\thg}\, =\, 0,  
\label{7.1} 
\ee 
with 
\be 
\og^2\, =\, (u^{\vf})^2\, =\, \frac{\ell^2}{R^4}.   
\label{7.3} 
\ee 
We see that indeed either $n^{\thg} = 0$, or else $n^{\thg}$ 
oscillates at frequency $f = \og/2\pi$, as the orbit is tilted 
with respect to the equatorial plane. 

Next we solve for the deviations from the circular orbit 
within the plane. For the components $(n^t, n^r, n^{\vf})$ 
the deviation equations take the general form 
\be 
\ddot{n}^i\, +\, \gam^i_{\;j} \dot{n}^j\, +\, m^i_{\;j} n^j\, 
 =\, 0, \hspace{2em} i,j = (t,r,\vf).   
\label{n.2} 
\ee 
The coefficients in this equations have the structure 
\be 
\mbox{\boldmath{$\gam$}}\, =\, \lh \ba{ccc} 
       0 & \gam^t_{\;r} & 0 \\ 
       \gam^r_{\;t} & 0 & \gam^r_{\;\vf} \\ 
       0 & \gam^{\vf}_{\;r} & 0 \ea \rh, \hspace{2em} 
{\bf m}\, =\, \lh \ba{ccc} 
       0 & 0 & 0 \\ 

       0 & m_r & 0 \\ 
       0 & 0 & 0 \ea \rh. 
\label{n.3}
\ee 
The precise form of the matrix elements follows from 
eq.(\ref{2.7}) or (\ref{nce}); they will be given shortly. 
Now our treatment applies to orbits deviating little from a 
circular orbit, which still decribe bound states. Therefore 
$r = R + \Del r$, with $\Del r = n^r \Del \lb$, must remain 
bounded, e.g.\ we look for oscillating solutions of 
(\ref{n.2}). From eqs.\ (\ref{n.2}), (\ref{n.3}) it then 
follows that with standard initial conditions (clocks and 
angles synchronized in the initial apastron: $n^t(0) = 
n^{\vf}(0) = 0, n^r(0) =$ maximal), the solutions take the 
form  
\be 
n^t = n^t_0 \sin \omega_1 \tau, \hspace{1em}
n^r = n^r_0 \cos \omega_1 \tau, \hspace{1em}
n^{\varphi} = n^{\varphi}_0 \sin \omega_1 \tau.
\label{n.4} 
\ee 
With this Ansatz, eq.(\ref{n.2}) can be written more explicitly
as 
\be 
T^i_{\;j}\, n^j_0\, =\, 0, 
\label{n.4.1}
\ee
with the matrix ${\bf T}$ defined by 
\be 
\ba{l} 
{\bf T}\, = \\ 
~\lh \ba{ccc}
  - \og_1^2 &
  - \og_1 \lh 2 \Gam_{rt}^{\;\;\;\;t} u^t - \frac{q}{m} F^t_{\;\;r} 
  \rh
  & 0 \\
  & & \\
  \og_1 \lh 2 \Gam_{tt}^{\;\;\;\;r} u^t - \frac{q}{m} F^r_{\;\;t} 
  \rh & \pl_r \Gam_{tt}^{\;\;\;\;r} (u^t)^2 + 
  \pl_r \Gam_{\vf\vf}^{\;\;\;\;r} (u^{\vf})^2 & 
  2 \og_1 \Gam_{\vf\vf}^{\;\;\;\;r} u^{\vf} \\ 
  & - \frac{q}{m} F^r_{\;t,r} u^t - \og_1^2 & \\
  & & \\ 
  0 & - 2 \og_1 \Gam_{r\vf}^{\;\;\;\;\vf} u^{\vf} & - \og_1^2 
  \ea \rh.
\ea 
\label{n.5}
\ee 
The solutions for the frequency $\og_1^2$ follow by solving the 
characteristic equation $\det {\bf T} = 0$:
\be 
\og_1^4\, \lh C + \frac{q}{m} \ag + \og_1^2 \rh\, =\, 0, 
\label{n.6}
\ee 
with the coefficients defined by 
\be 
\ba{l} 
C = \lh - \pl_r \Gam_{tt}^{\;\;\;\;r} + 4 
  \Gam_{rt}^{\;\;\;\;t} \Gam_{tt}^{\;\;\;\;r} \rh\, (u^t)^2 + 
  \lh - \pl_r \Gam_{\vf\vf}^{\;\;\;\;r} + 4 
  \Gam_{r\vf}^{\;\;\;\;\vf} \Gam_{\vf\vf}^{\;\;\;\;r} \rh\, 
  (u^\vf)^2, \\
  \\ 
\dsp{ \frac{q}{m}\, \ag = \frac{q}{m}\, \lh F^r_{\;t,r} 
  - 4 \Gam_{rt}^{\;\;\;\;t} F^r_{\;\;t} \rh u^t + 
  \frac{q^2}{m^2}\, F^r_{\;\;t} F^t_{\;\;r}. }
 \ea
\label{n.7}
\ee 
Disregarding the trivial solution $\og_1^2 = 0$, which occurs
with double degeneracy, the solution of (\ref{n.6}) of interest
to us is
\be 
\og_1^2 = - \lh C + \frac{q}{m} \ag \rh. 
\label{n.8}
\ee 
We compute the values of $\ag$ to zeroth, and $C$ to first 
order in $q/m$. The connection coefficients 
$\Gam_{ij}^{\;\;\;\;k}$ have been collected in appendix 
(\ref{A}.a). The electromagnetic field strength is given 
in eq.(\ref{6.3}). Finally, eqs.\ (\ref{6.4}) and (\ref{6.5}),
toegether with (\ref{6.9}), imply for the velocities 
\be  
(u^t)^2 = \frac{R^2 + \ell^2}{R^2 - 2MR + Q^2}
 \approx \frac{\ell^2}{MR - Q^2} \lh 1 + 
 \frac{qQ}{4\pi mM} + ... \rh, 
\label{n.9} 
\ee 
where we have neglected terms of higher order in the small 
parameters $(M/R, Q/M, q/m)$; and 
\be 
(u^{\vf})^2 = \og^2 = \frac{\ell^2}{R^4}. 
\label{n.10}
\ee 
As a result of the various substitutions one obtains  
\be 
\ba{rcl} 
C & = & \dsp{ 
  - \og^2 \left[1 - \frac{6M}{R} + 
  \frac{Q^2}{MR} - \frac{2qQ}{4\pi mM} + ... \right], }\\ 
 & & \\ 
\dsp{ \frac{q}{m} \ag } & = & \dsp{ - 
 \frac{2qQ}{4\pi mM}\, \og^2 + ... }
\ea 
\label{n.11}
\ee 
Observe that all terms of order $qQ/4\pi mM$ cancel. 
Indeed, deviations in the orbital period are not expected 
from the Coulomb interaction, but only from relativistic 
corrections to the Coulomb interaction, which are of higher 
order in $(M/R, Q/M, q/m)$. 

The solution of the characteristic equation 
in this approximation thus becomes 
by 
\be 
\og_1^2\, =\, \og^2\, \lh 1 - \frac{6M}{R} 
 + \frac{Q^2}{MR} + ... \rh,
\label{n.12}
\ee 
which gives for the frequency 
\be 
\og_1\, =\, \og\, \lh 1  - \frac{3M}{R} 
 + \frac{Q^2}{2MR} + ... \rh.
\label{n.13}
\ee 
To complete the solutions (\ref{n.4}), eq.(\ref{2.8.1}) 
establishes a relation between the components $n^{\mu}$, 
which for the solutions (\ref{n.4}) becomes 
\be 
\lh \ve - \frac{qQ}{4\pi mR} \rh n^t_0\, -\, \ell n^{\vf}_0\, 
 =\, \frac{qQ}{4\pi m\og_1 R^2}\, u^t n^r_0. 
\label{n.14}
\ee 
This guarantees that the perturbed orbit is again a solution 
of the equations of motion. We can interpret this in terms of 
observable co-ordinate differences by writing $\Del x^{\mu} = 
n^{\mu} \Del \lb$. 

We finish with some physical observations relating to our 
solution (\ref{n.4}) with frequency (\ref{n.13}). First, 
note that there are different solutions $\og_1$ for different 
values of the test charge $q$: there is the frequency $\og_0$ 
for neutral test particles $(q = 0)$, a larger one $\og_+$ 
for test and central charge of equal sign, and a smaller one 
$\og_-$ for test and central charge of opposite sign.  

If we compare our approximate solution with the parametrized 
exact solution (\ref{6.7.2}): 
\be 
r(\vf)\, \approx\, R\, +\, \Del r\, \approx\, 
 R\, -\, eR \cos y(\vf), 
\label{n.15}
\ee 
we find, using the initial conditions in the apastron 
$(y = - \pi)$, that to first order 
\be 
e\, =\, \frac{(\Del r)_0}{R}\, =\, \frac{n^r_0\, \Del \lb}{R}. 
\label{n.16}
\ee 
Thus our solution describes an approximate ellipse (Kepler 
orbit), with eccentricity $e$. 

We can also calculate the precession of the periastron. 
Observe that as the orbit reaches its extremal radius, 
the shifts in orbital angle and time vanish: $n^r_{extr} 
= \pm\, n^r_0$ at $\og_1 \tau = k \pi$, with $k$ integer; 
therefore $n^t_{extr} = n^{\vf}_{extr} = 0$. Now we can 
calculate the period between two periastra of the perturbed 
orbit: 
\be 
\ba{lll}
T & = & \dsp{ \int_0^{2\pi/\og_1} d\tau\, \frac{dt}{d\tau}\, 
 =\, \int_0^{2\pi/\og_1} d\tau\, \lh u^t + \dot{n}^t 
 \Del \lb \rh }\\ 
 & & \\
 & = & \dsp{ \frac{2\pi}{\og_1}\, u^t\, +\, \left[ n^t 
 \Del \lb \right]_0^{2\pi/\og_1}\, =\, \frac{2\pi}{\og_1}\, 
 u^t.} 
\ea 
\label{n.17}
\ee 
Here $u^t$ denotes the constant value (\ref{n.9}) of $dt/d\tau$ 
for the circular reference orbit. At this time the values 
of the angular co-ordinate of the perturbed and circular 
reference orbits coincide. The angular direction of the 
periastron can therefore be calculated by calculating the 
angular co-ordinate of the circular orbit at time $T$. The 
angular shift of the periastron per period is then 
\be 
\del \vf\, =\, \bar{\vf}(\bar{t} = T)\, -\, 
 \bar{\vf}(0)\, -\, 2\pi, 
\label{n.18}
\ee 
where we have used overlines to denote the co-ordinates 
of the circular orbit. Now as $u^{\vf}$ is conserved, 
we have 
\be 
\bar{\vf}(T)\, -\, \bar{\vf}(0)\, =\, 
 \frac{\ell}{R^2} \frac{T}{u^t}\, =\, 
 \frac{\og T}{u^t}. 
\label{n.19}
\ee 
Substitution of this result into eq.(\ref{n.18}), 
using (\ref{n.13}), gives
\be 
\del \vf\, =\, 2\pi \lh \frac{\og}{\og_1} - 1 \rh\, 
 \approx\, 2\pi \lh \frac{3M}{R} - \frac{Q^2}{2MR} 
 + ... \rh. 
\label{n.20}
\ee 
This is in complete agreement with eq.(\ref{6.7.7}). 

\section{Conclusion}

The method of investigation of particle motions close to
the exact solutions already known, can be used in other
exact solutions of Einstein-Maxwell system. We believe that
it is worthwhile to pursue the efforts in this direction,
in particular because this method introduces the small
parameter (the norm of the deviation vector) different from
the small parameters used by many authors in search for
approximate solutions of the theory, such as e.g. the
ratio $M/R$ or $Q/R$. It is also well-suited for computer
calculations.

\vskip 0.5cm
\indent
{\bf Acknowledgement}
\vskip 0.3cm
One of us (R.K.) would like to thank Dr. Christian Klein
for many useful and enlightening discussions.

\newpage 
\appendix

\section{Appendix: Connections and curvatures for 
Reissner-Nordstr{\o}m geometry \label{A}} 
 
In this appendix we collect the expressions for the components 
of the connections and Riemann curvature used in the main body of 
the paper.  
\newline
\noindent
{\em a.\ Connections.} From the line-element (\ref{6.1}) one 
derives the following expressions for the connection co-efficients: 
\be 
\ba{l} 
\dsp{ \Gam_{rt}^{\;\;\;\;t} = - \Gam_{rr}^{\;\;\;\;r} = 
 \frac{Mr - Q^2}{r \lh r^2 - 2Mr + Q^2 \rh}, } \\ 
 \\ 
\dsp{ \Gam_{tt}^{\;\;\;\;r} = \frac{1}{r^5}\, \lh Mr - Q^2 \rh 
 \lh r^2 - 2Mr + Q^2 \rh, } \\ 
 \\ 
\dsp{ \Gam_{\vf\vf}^{\;\;\;\;r} = \sin^2 \thg\, 
 \Gam_{\thg\thg}^{\;\;\;\;r} = -\, \frac{\sin^2 \thg}{r}\, \lh 
 r^2 - 2Mr + Q^2 \rh, }\\ 
 \\ 
\dsp{ \Gam_{r\thg}^{\;\;\;\;\thg} = \Gam_{r\vf}^{\;\;\;\;\vf} = 
  \frac{1}{r}, }\\ 
 \\ 
\dsp{ \Gam_{\thg\vf}^{\;\;\;\;\vf} = \frac{\cos \thg}{\sin \thg}, 
  \hspace{3em} \Gam_{\vf\vf}^{\;\;\;\;\thg} = - \sin \thg\, 
  \cos \thg.} 
\ea 
\label{a.1} 
\ee 
 
\nit 
{\em b.\ Curvature components.} The corresponding curvature 
two-form components $R_{\mu\nu} = \frac{1}{2}\, R_{\kg \lb \mu\nu} 
dx^{\kg} \wedge dx^{\lb}$ are: 
\be  
\ba{l} 
\dsp{ R_{tr} = \frac{1}{r^4}\, \lh 2Mr - 3Q^2 \rh\, dt \wedge dr, 
}\\ 
 \\ 
\dsp{ R_{t\thg} = -\, \frac{1}{r^4}\, \lh Mr - Q^2 \rh\, 
 \lh r^2  - 2Mr + Q^2 \rh\, dt \wedge d\thg, }\\ 
 \\ 
\dsp{ R_{t\vf} = -\, \frac{1}{r^4}\, \lh Mr - Q^2 \rh\, 
 \lh r^2  - 2Mr + Q^2 \rh\, \sin^2 \thg\, dt \wedge d\vf, } 
\ea 
\label{a.2} 
\ee 
and  
\be  
\ba{l} 
\dsp{ R_{r\thg} = \frac{Mr - Q^2}{r^2 - 2Mr + Q^2}\, dr \wedge 
 d\thg, }\\ 
 \\ 
\dsp{ R_{r\vf} = \frac{Mr - Q^2}{r^2 - 2Mr + Q^2}\, \sin^2 \thg\, 
 dr \wedge d\vf,  }\\ 
 \\ 
\dsp{ R_{\thg\vf} = -\, \lh 2Mr - Q^2 \rh\, \sin^2 \thg\, 
 d\thg \wedge d\vf. } 
\ea 
\label{a.3} 
\ee

\end{document}